\documentclass{jpp}

\usepackage{graphicx}
\usepackage[utf8]{inputenc}
\usepackage[T1]{fontenc}
\usepackage{natbib}
\usepackage[colorlinks,urlcolor=blue,citecolor=blue,linkcolor=blue]{hyperref}
\usepackage{amsmath}
\usepackage{subcaption}
\usepackage[export]{adjustbox}
\usepackage{gensymb}

\ifCUPmtlplainloaded \else
  \checkfont{eurm10}
  \iffontfound
    \IfFileExists{upmath.sty}
      {\typeout{^^JFound AMS Euler Roman fonts on the system,
                   using the 'upmath' package.^^J}%
       \usepackage{upmath}}
      {\typeout{^^JFound AMS Euler Roman fonts on the system, but you
                   dont seem to have the}%
       \typeout{'upmath' package installed. JPP.cls can take advantage
                 of these fonts, if you use 'upmath' package.^^J}%
      }
  \else
  \fi
\fi

\ifCUPmtlplainloaded \else
  \checkfont{msam10}
  \iffontfound
    \IfFileExists{amssymb.sty}
      {\typeout{^^JFound AMS Symbol fonts on the system, using the
                'amssymb' package.^^J}%
       \usepackage{amssymb}%

      }{}
  \fi
\fi

\ifCUPmtlplainloaded \else
  \IfFileExists{amsbsy.sty}
    {\typeout{^^JFound the 'amsbsy' package on the system, using it.^^J}%
     \usepackage{amsbsy}}
    {\providecommand\boldsymbol[1]{\mbox{\boldmath $##1$}}}
\fi

\def\comp{\,c/\omega_{\rm p}}

\newcommand{\fig}[1]{Fig.~\ref{fig:#1}}

\newcommand{\be}{\begin{equation}} 
\newcommand{\ee}{\end{equation}}

\newcommand{\ba}{\begin{eqnarray}}
\newcommand{\ea}{\end{eqnarray}}

\newcommand{\Alfven}{Alfv\'{e}n }

\newcommand{\Bf}{{magnetic field}}
\newcommand{\Bfs}{{magnetic fields}}

\newcommand{\NS}{neutron star}
\newcommand{\NSs}{{neutron stars}}

\newcommand{\ms}{magnetosphere}
\newcommand{\mss}{magnetospheres}

\newcommand{\aap}{    {\it Astron. Astrophys.}}
\newcommand{\aaps}{   {\it Astron. Astrophys. Suppl.}}
\newcommand{\aapr}{   {\it Astron. Astrophys. Rev.}}

\newcommand{\aj}{     {\it Astron. J.}}
\newcommand{\apj}{    {\it Astrophys. J.}}
\newcommand{\apjl}{    {\it Astrophys. J. Lett.}}
\newcommand{\apss}{   {\it Astrophys. Space Sci.}}

\newcommand{\fcp}{    {\it Fundamenals Cosm. Phys.}}

\newcommand{\grl}{    {\it Geophys. Res. Lett.}}

\newcommand{\jgr}{    {\it J. Geophys. Res.}}
\newcommand{\mnras}{  {\it Mon. Not. Roy. Astron. Soc.}}
\newcommand{\nat}{    {\it Nature}}
\newcommand{\pasp}{   {\it Pub. Astron. Soc. Pac.}}
\newcommand{\pasj}{   {\it Pub. Astron. Soc. Japan}}
\newcommand{\prd}{    {\it Phys. Rev. D}}
\newcommand{\pre}{    {\it Phys. Rev. E}}
\newcommand{\solphys}{{\it Solar Phys.}}
\newcommand{\sovast}{ {\it Sov. Astron.}}
\newcommand{\ssr}{    {\it Space Sci. Rev.}}

\newcommand{\na}{    {\it Nature}}
\def\physrep{Phys.~Rep.}      
\def\araa{ARA\&A}             

\title[Mehta et al.]{Tilting Instability of Magnetically Confined Spheromaks}

\author{Riddhi Mehta\aff{1}
  \corresp{\email{mehta74@purdue.edu}},
  Maxim Barkov\aff{1,2,3}
  \corresp{\email{mbarkov@purdue.edu}},
  Lorenzo Sironi\aff{4}
  \corresp{\email{lsironi@astro.columbia.edu}}
 \and Maxim Lyutikov\aff{1}
 \corresp{\email{lyutikov@purdue.edu}}}

\affiliation{\aff{1} Department of Physics and Astronomy, Purdue University, 525 Northwestern Avenue, West Lafayette, IN, 47907-2036, USA
\aff{2} Astrophysical Big Bang Laboratory, RIKEN, 351-0198 Saitama, Japan
\aff{3} Institute of Astronomy, Russian Academy of Sciences, Pyatnitskaya 48, 119017, Moscow, Russian Federation
\aff{4} Department of Astronomy, Columbia University, 550 West 120th street, New York, NY 10027, USA}

\begin{document}

\maketitle
\begin{abstract}
We consider the tilting instability of a magnetically confined spheromak using 3D MHD and relativistic PIC calculations with an application to astrophysical plasmas, specifically those occurring in magnetar magnetospheres. The instability is driven by the  counter alignment of the spheromak's intrinsic  magnetic dipole with  the external \Bf. Initially the spheromak rotates - tilts - trying to lower its magnetic potential energy. As a result a current sheet forms between the internal \Bf\ of a spheromak and the confining field. Magnetic reconnection sets in; this  leads to the   annihilation of the  newly counter-aligned  magnetic flux  of the spheromak. This occurs on few \Alfven  time scales. In the case of higher order (second order) spheromak, the internal core is first pushed out of the envelope, resulting in formation of  two nearly independent tilting spheromaks. Thus, the magnetically twisted outer shell cannot stabilize the inner core. During  dissipation, helicity of the initial  spheromak is carried away by torsional \Alfven waves, violating the assumptions of the Taylor relaxation  theorem. In applications to magnetars' giant flares, 
fast development of tilting  instabilities, and no stabilization  of the higher order spheromaks, make it unlikely that trapped spheromaks are responsible for the tail emission lasting hundreds of seconds.
\end{abstract}


    \section{Introduction}
    
    Relaxation of magnetized plasma is a fundamental problem in laboratory and space plasma physics \citep{Woltier:1958,1974PhRvL..33.1139T,2000mare.book.....P}. 
    In this work we are particularly interested in the relaxation processes in highly magnetized astrophysical plasmas, where the magnetic field controls the overall dynamics
of the plasma, and the dissipation of magnetic energy may power the observed high-energy
emission.  The most relevant  astrophysical settings include magnetars \citep[strongly magnetized \NSs\ possessing
super-strong \Bfs][]{Duncan2001,2017ARA&A..55..261K}, pulsars and pulsar wind nebulae \cite[(PWNe)][]{2006ARA&A..44...17G}, jets of Active Galactic
Nuclei (AGNs), and Gamma-Ray Bursters \citep[(GRBs)][]{2006NJPh....8..119L}.  All these objects are efficient emitters of X-rays
and $\gamma$-rays, and in the past two decades they have been the subjects of 
intensive observational studies via a number of  successful high-energy
satellites.  These objects seem to 
share one important property -  they include relativistic
magnetized plasmas, and often the plasma is magnetically dominated, \textit{i.e.},  
the energy density of this plasma is mostly contributed not by the rest mass  energy of matter, but by the energy of the magnetic field. This is dramatically different  
from laboratory plasmas, magnetospheres of planets and interplanetary 
plasma. This extreme regime can only be probed (though, indirectly) via observations of relativistic astrophysical sources, by unveiling the 
imprint left by the magnetic field dissipation on the observed emission. 

In addition to high (relativistic) magnetization, astrophysical plasmas differ from laboratory ones by the absence of pre-arranged conducting walls. This has important implications for stability, and the applicability  of the Taylor relaxation principle as we discuss below.

\section{Spheromaks and MHD relaxation}
    
    Particularly important are static equilibria when MHD equations demand
    \be
    \nabla p = {\bf J} \times  {\bf B} 
    \ee
    where $p$ is plasma pressure, and ${\bf J} $ and $ {\bf B}$ are current density and \Bf. For magnetically dominated regimes the pressure gradient is negligible, and plasma equilibrium becomes a force-free equilibrium \citep{ChandrasekharKendall57}
    \be
    \boldsymbol{J}\times\boldsymbol{B} \approx 0
    \label{jcrosB}
    \ee
   Of particular importance is the Taylor state, where $ {\bf J}  \propto  \nabla \times {\bf B} = \lambda    {\bf B}$ with spatially constant $ \lambda $. 
An initially turbulent plasma is expected to spontaneously relax (or self-organize) to a simple, well-defined  Taylor state.
      In a finite volume the system reaches a state with the smallest possible $ \lambda $ (largest scale configurations). In cylindrical geometry the corresponding configurations - Lundquist states   \citep{Lundquist50} - are indeed the endpoints of relaxation \citep{1987RPPh...50..115K}. Importantly, Lundquist states are, in a sense, connected to walls - they extend infinitely along the symmetry axis.

    In spherical geometry the force-free configurations with constant $ \lambda $ are  called  spheromaks \citep{Rosenbluth1979,Bellan2000}.
      Spheromaks have a number of features that make them  useful as basic plasma structures,   building blocks of plasma models. 
   First, spheromaks are not connected to any confining wall such as that of a laboratory vessel or to coils and hence represent a ``pure'' kind of plasma configuration that could be achieved by internal plasma relaxation. Internally spheromaks are simply connected (not  topological tori).
   Second, they represent a    relaxed (Taylor) state  - one might expect that   a turbulent plasma would spontaneously relax (or self-organize) to a simple state resembling a spheromak. 
   
 Astrophysical plasmas like those found in magnetar magnetospheres \citep{Duncan2001,Masada2010,Lyutikov2003,Komissarov2006} are likely to evolve in to a force-free configuration, effectively confined through the creation of a system of nested poloidal flux surfaces.  Given the appropriate initial conditions, spheromaks can form spontaneously due to plasma instabilities and hence can be hypothesized to form in an astrophysical environment. For example,  it was suggested that spontaneous instabilities arising in plasmas can lead to a spheromak configuration which suggests that such configurations should occur in nature. Indeed, the magnetically confined fireball picture has been invoked to explain coronal mass ejections arising in solar flares \citep{Ivanov1985,Masada2010,2011SoPh..270..537L} and high energy flaring/bursting activity of magnetars \citep{Duncan2001,Lyutikov2003,2008MNRAS.387.1735M}.

 In this paper we are mostly interested in astrophysical applications, particularly  in the high magnetization regime. First, in that case the effects of finite gyroradius are not important. For example, in the magnetospheres of magnetars the magnetic  field is of the order of the quantum field, so that even at relativistic temperatures the gyro-radius  is only $\sim  ~ 10^{-11}$ cm, many orders of magnitude smaller than the expected overall size of $\sim 10^6$ cm.  Thus, astrophysical configurations we are interested in are well in the MHD regime. Secondly,  stability of spheromaks and Field-Reversed Configurations (FRC)  in laboratory setting depends on the arrangement of confining conducting walls  \citep{Rosenbluth1979,Hayashi1983,2000PhPl....7.4996B,2006PhPl...13e6115B}. In contrast, astrophysical configurations are generally expected to be less affected by the presence of conducting walls. Spheromaks also present  a simple analytically tractable configuration, as opposed to FRC configurations where initial state has to be calculated  numerically.

 
 In contrast to the cylindrical  Lundquist case, the 3D magnetically confined  basic spheromaks are unstable in the absence of conducting walls \citep{Rosenbluth1979,Hayashi1983}.
The basic reason for instability is that the magnetic dipole moment of  a trapped spheromak is anti-aligned with an external  magnetic field. As  a result, a magnetically confined spheromak is intrinsically unstable and would prefer to tilt to lower its magnetic potential energy.  A number of authors considered stabilizing effects of conducting magnets on the evolution of the spheromak  \citep{Bondeson1981,Finn1981,2001PhPl....8.1267B}; see \cite{1994PPCF...36..945J} for review of spheromak research. \\
 
  In this paper, we reanalyze the structure and time evolution of magnetically confined spheromaks using 3D MHD and PIC simulations with an application to astrophysical plasmas occurring in magnetar magnetospheres. Previously, 
 reconnection and particle acceleration due to current-driven instabilities in Newtonian, initially force-free plasmas in 2.5D and 3D scenarios using high-resolution simulations both with a fixed grid and with adaptive mesh refinement is studied extensively in \cite{Ripperda2017}. In 2.5D, the two parallel repelling current channels in an initially force-free equilibrium are first subject to a linear instability consisting of an antiparallel displacement and thereafter undergo a rotation and twisting motion. They quantify the growth rate of this tilting instability by a linear growth phase in the bulk kinetic energy during which reconnection of magnetic field lines causes the formation of nearly-singular current sheets and secondary islands leading to particle acceleration. Our 3D MHD simulation (\S\hyperref[sec:2.3]{3.3} and \fig{sp1z} \& \fig{toroidal}) of the force-free spheromak clearly displays the onset of a similar tilt instability and twisting motion which leads to magnetic reconnection at the boundaries between the spheromak and the external field, causing the spheromak to eventually dissipate.

\section{Spheromak in External Magnetic Field}
\label{sec:2}

\subsection{Basic Spheromak}
\label{sec:AM}

Let us first briefly recall the structure of magnetically confined spheromaks. 
In the  Grad-Shafranov formalism \citep{1967PhFl...10..137G,Shafranov} 
  the magnetic field can be represented by a scalar flux function $\psi$ in spherical coordinates 
\begin{equation}
    \boldsymbol{B}=\nabla\psi\times\nabla\phi+\lambda\psi\nabla\phi
    \label{axisB}
\end{equation}
where $\phi$ is the toroidal coordinate. An axisymmetric solution of Eq. (\ref{axisB}) within a sphere of radius $r_0$ and constant $\lambda$  is a spheromak \citep{Rosenbluth1979,Bellan2000}.

Using Eq. (\ref{axisB}) and condition for the Taylor state, the Grad-Shafranov equation (GSE) of axisymmetric force-free toroidal plasma equilibrium can be represented in spherical coordinates \citep{Tsui2008}
\begin{equation}
    r^2\frac{\partial^2\psi}{\partial r^2}+\sin\theta\frac{\partial}{\partial\theta}\left(\frac{1}{\sin\theta}\frac{\partial\psi}{\partial\theta}\right)+(\lambda r)^2\psi=0
    \label{GSE}
\end{equation}
Eq. (\ref{GSE}) can be solved using  separation of variables inside and outside the spheromak.

Inside the spheromak, magnetic field components are
\begin{equation}
\left. \begin{array}{l}
\displaystyle 
{B}_{r} = 2A_0\frac{\lambda}{r}j_1(\lambda r)\cos\theta\\[16pt]
\displaystyle 
{B}_{\theta} = -A_0\frac{\lambda}{r}\frac{\partial}{\partial r}(r j_1(\lambda r))\sin\theta\\[16pt]
\displaystyle 
{B}_{\phi} = A_0{\lambda}^2j_1(\lambda r)\sin\theta
\end{array} \right\}
\label{Bin}
\end{equation}\\
where, $j_1(\lambda r)$ is  spherical Bessel function of the first kind.

The radial and toroidal components of magnetic field vanish on the surface of spheromak which corresponds to $j_1(\lambda r) = 0$ at $r=r_0$. This gives the smallest allowed $\lambda$ corresponding to the lowest energy Taylor state
\begin{equation}
  \lambda \approx 4.493/{r_0} 
  \label{taylorstate}
\end{equation}

Outside the spheromak, magnetic field is
\be
    \boldsymbol{B_{\rm ex}} = \left(B_0\cos\theta - B_0\cos\theta \frac{r_0^3}{r^3},-B_0\sin\theta - B_0\sin\theta\frac{r_0^3}{2r^3},0\right)
\end{equation}
where, magnetic field at very large distances asymptotes to a uniform field $B_0 \hat{z}$. Since, magnetic field at the surface of the spheromak is continuous, the constant $A_0$ can be related to the external  magnetic field $B_0$ 
\begin{equation}
    A_0 \approx -0.342B_0r_0^2
\ee

\subsection{Tilt Instability of Spheromak in External Magnetic Field}
\label{sec:tilt}

The basic magnetically confined spheromak is unstable. The easiest  way to see this is to note that a spheromak can be approximated as a magnetic dipole $\boldsymbol{\mu}$  embedded in an external  magnetic field

\begin{equation}
    \boldsymbol{\mu} = \frac{-B_0r_0^3}{2}\hat{z}  
    \label{mu}
\end{equation}
Eq. (\ref{mu}) shows that the magnetic moment of a spheromak is anti-aligned with the external magnetic field  and hence subject to tilt. Tilt instability of spheromak has been explored extensively by \cite{Bellan2000} and \cite{Jardin1986}, both of which serve to validate the arguments made in this paper. 

In \cite{Bellan2000} the spheromak is described as a small magnet between two large magnets oriented anti-parallel to large external magnets hence unstable to tilting. The flipping of a spheromak by {180\degree} to lower its potential energy, however, causes the external field to be such as to enhance rather than balance the spheromak hoop force. Equilibrium is quickly lost and the spheromak will explode outwards at Alfv{\'e}n velocity. Our 3D MHD simulations of \S\hyperref[sec:2.3]{3.3} show this dissipation of spheromak after undergoing tilt instability and aid us to estimate the dissipation timescale in units of Alfv{\'e}nic crossing time. 

In \cite{Jardin1986}, a spheromak is described simply as a rigid current carrying ring and its various rigid instabilities like tilting, shifting and vertical motions are discussed as modes which get activated depending on the value of the magnetic field index $n=-(r/B_0)(\partial{B_0}/\partial{r})$ where $B_0$ is the magnitude of the external vertical magnetic field. The tilting mode is unstable for $n<1$. For laboratory spheromak experiments, the growth rate of these instabilities which would eventually cause the spheromak to dissipate, is estimated to be $1-10\mu$s. We estimate such a timescale for astrophysical plasmas using results of our 3D MHD simulations.


\subsection{3D MHD Simulations of  Tilting Instability}
\label{sec:2.3}
\subsubsection{Numerical Setup}

We perform 3D MHD simulations of the lowest energy Taylor state as described by Eqs.(\ref{Bin}) and (\ref{taylorstate}) as well as the 2-root spheromak with constant--density uniformly magnetized plasma to explore their time evolution and test their stability. The simulations were performed using a three dimensional (3D) geometry in Cartesian coordinates using the \textit{PLUTO} code\footnote{http://plutocode.ph.unito.it/index.html} \citep{Mignone2007}. \textit{PLUTO} is a modular Godunov-type code entirely written in C, intended mainly for astrophysical applications and high Mach number flows in multiple spatial dimensions and designed to integrate a general system of conservation laws

\begin{equation}
    \frac{\partial \boldsymbol{U}}{\partial t} = -\nabla.\bf{T}(\boldsymbol{U}) + \boldsymbol{S}(\boldsymbol{U})
    \label{tensor}
\end{equation}
$\boldsymbol{U}$ is the vector of conservative variables and $\bf{T}(\boldsymbol{U})$ is the matrix of fluxes associated with those variables. For our ideal MHD setup, no source terms are used and $\boldsymbol{U}$ and $\bf{T}$ are

\begin{equation}
\boldsymbol{U} = \begin{pmatrix} \rho\\ \boldsymbol{m}\\ \boldsymbol{B}\\ E  \end{pmatrix},
\bf{T}(\boldsymbol{U}) = \begin{bmatrix} \rho \boldsymbol{v} \\ \boldsymbol{mv} - \boldsymbol{BB} + p_t \bf{I} \\ \boldsymbol{vB - Bv} \\ (E + p_t)\boldsymbol{v} - \boldsymbol{(v.B)B} \end{bmatrix}^{\textit{T}}
\label{U&T}
\end{equation}
$\rho, \boldsymbol{v}$ and $p$ are density, velocity and thermal pressure. $\boldsymbol{m} = \rho \boldsymbol{v}$, $\boldsymbol{B}$ is the magnetic field and $p_t = p + |{\boldsymbol{B}}|^{2}/2$ is the total (thermal + magnetic) pressure, respectively. Magnetic field evolution is complemented by the additional constraint $\nabla.\boldsymbol{B}=0$. Total energy density $E$

\begin{equation}
    E = \frac{p}{\Gamma - 1} + \frac{1}{2}\left(\frac{|\boldsymbol{m}|^{2}}{\rho} + |{\boldsymbol{B}}|^{2}\right)
    \label{energy_density}
\end{equation}
along with an isothermal equation of state $p = c_s^2 \rho$ provides the closure. $\Gamma$ and $c_s$ are the polytropic index and isothermal sound speed, respectively.  \textit{MP5\_FD} interpolation, a 3rd order Runge-Kutta approximation in time, and an HLL Riemann solver are used to solve the above ideal MHD equations.

The plasma has been approximated as an ideal, non-relativistic adiabatic gas, one particle species with polytropic index of 5/3. The size of the domain is $x \in [-2, 2]$ and $y \in [-2, 2]$, $z \in [-3.3, 3.3]$. To better resolve the evolution of spheromak, non-uniform resolution is used in the computational domain with total number of cells $N_{\rm X}=N_{\rm Y}$ = 312 and $N_{\rm Z}$ = 520. We also check that decreasing the resolution by a factor of two, that is, $N_{\rm X}=N_{\rm Y}$ = 156 and $N_{\rm Z}$ = 260, does not affect the simulation results. Convergence will be evident later in \fig{ins} and \fig{energy_1r}. Outflow boundary conditions are applied in all three directions.

In the simulation, values for constant external magnetic field $B_0$, radius of spheromak $r_0$ and plasma density $\rho$ were set to 0.3, 0.75 and 1 respectively. With a magnetization $\sigma=B_0^2/\rho=0.09$, Alfv{\'e}n speed $v_A$ is only mildly relativistic given by $v_A=B_0/\sqrt{\rho}=0.3$. Our motive here is to stress more on astrophysical applications, in particular magnetar’s magnetospheres, where Alfven velocity is expected to be relativistic. $B_0$, $r_0$ and $\rho$ are used to estimate a time scale of propagation of magnetic oscillations within the spheromak in terms of Alfv{\'e}nic crossing time $t_{A}=r_0/v_A = 2.5$. The timescale over which the spheromak disrupts is estimated later in units of $t_A$. Projections of total magnetic field and current density in the $xz$ plane are denoted by $\boldsymbol{B3D}$ and $\boldsymbol{J}$ respectively. All quantities are given in code units which are normalized $cgs$ values

\begin{equation}
    \rho = \frac{\rho_{cgs}}{\rho_n},
    v = \frac{v_{cgs}}{v_n},
    p = \frac{p_{cgs}}{\rho_n v_n^2},
    B = \frac{B_{cgs}}{\sqrt{4 \pi \rho_n v_n^2}}
     \label{code_units}
\end{equation}
$\rho$, $v$, $p$ and $B$ are density, velocity, pressure and magnetic field. Time is given in units of $t_n=L_n/v_n$. The normalization values used are $\rho_n = 1.67 \times 10^{-23} \text{gr/cm}^3$, $L_n = 3.1 \times 10^{18}$ cm and $v_n = 10^5$ cm/s. 

\subsubsection{Tilting Instability of Basic Spheromak }
\label{sec:taylor}

We perform two types of simulations: one with resolution $312\times312\times520$ (I) and another with resolution $156\times156\times260$ (II). The following discussion describes results for simulation (I). \fig{sp1z}, \fig{toroidal} and \fig{current} display time evolution of a basic magnetically confined  spheromak.  Shown are the  2D ($xz$ plane) slices of a 3D simulation. Vectors in \fig{sp1z} $\&$ \fig{toroidal} denote total magnetic field projected in the $xz$ plane and those in \fig{current} depict total current density projected in the $xz$ plane. Color bars in \fig{sp1z} show plasma density, those in \fig{toroidal} show toroidal magnetic field $B_y$ and those in \fig{current} show toroidal current density $J_y$. Starting from $t=0$, the spheromak is captured at subsequent time instants where significant changes to its morphology can be observed.

\begin{figure}
\centering
\includegraphics[width=\textwidth]{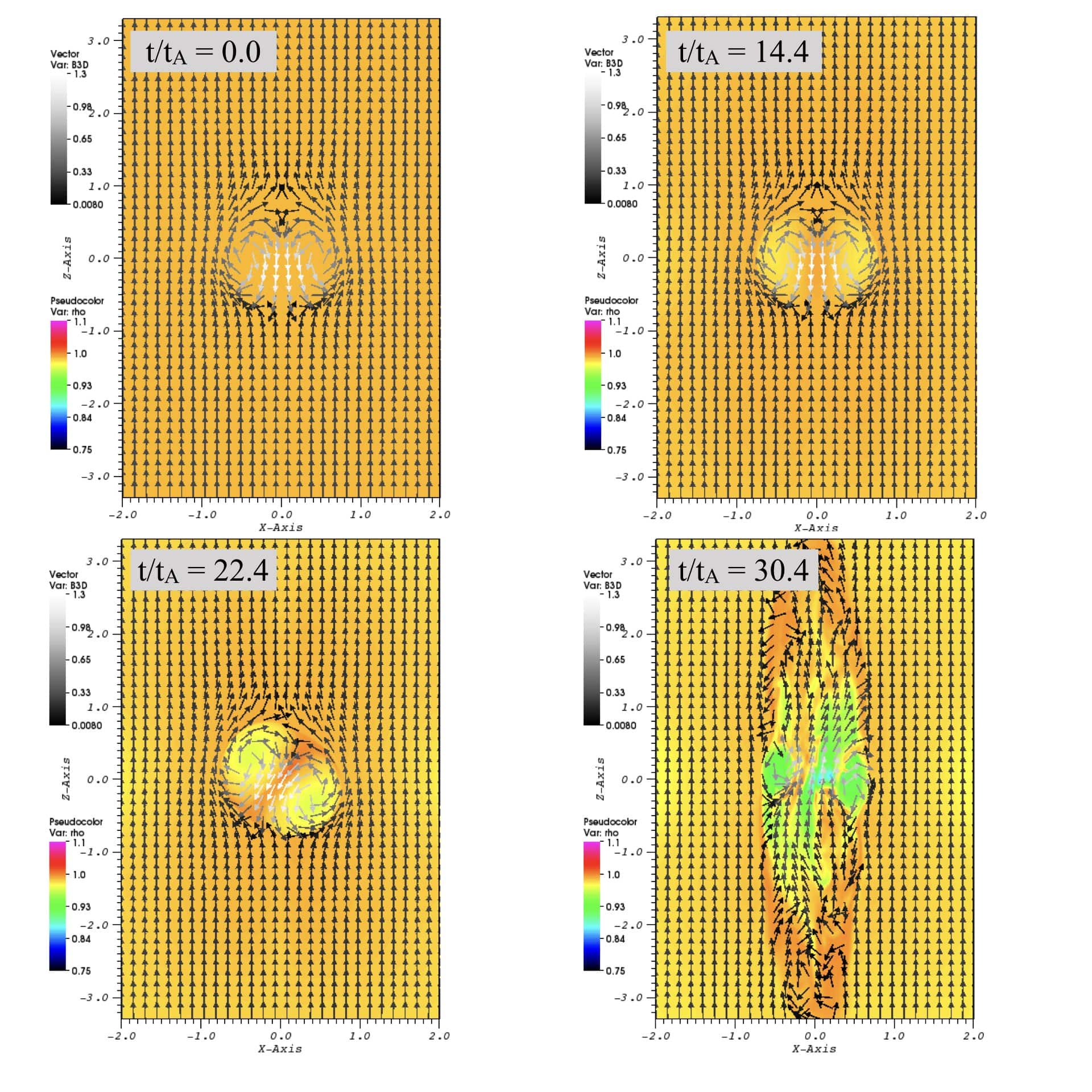}
\caption{Slice in the $xz$ plane of MHD simulation of lowest energy Taylor state. Times indicated in the panels are in units of the Alfv\'{e}nic crossing time $t_A=r_0/v_A$. Colors indicate plasma density while vectors depict $\boldsymbol{B3D}$.}
\label{fig:sp1z}
\end{figure}

\begin{figure}
\centering
\includegraphics[width=\textwidth]{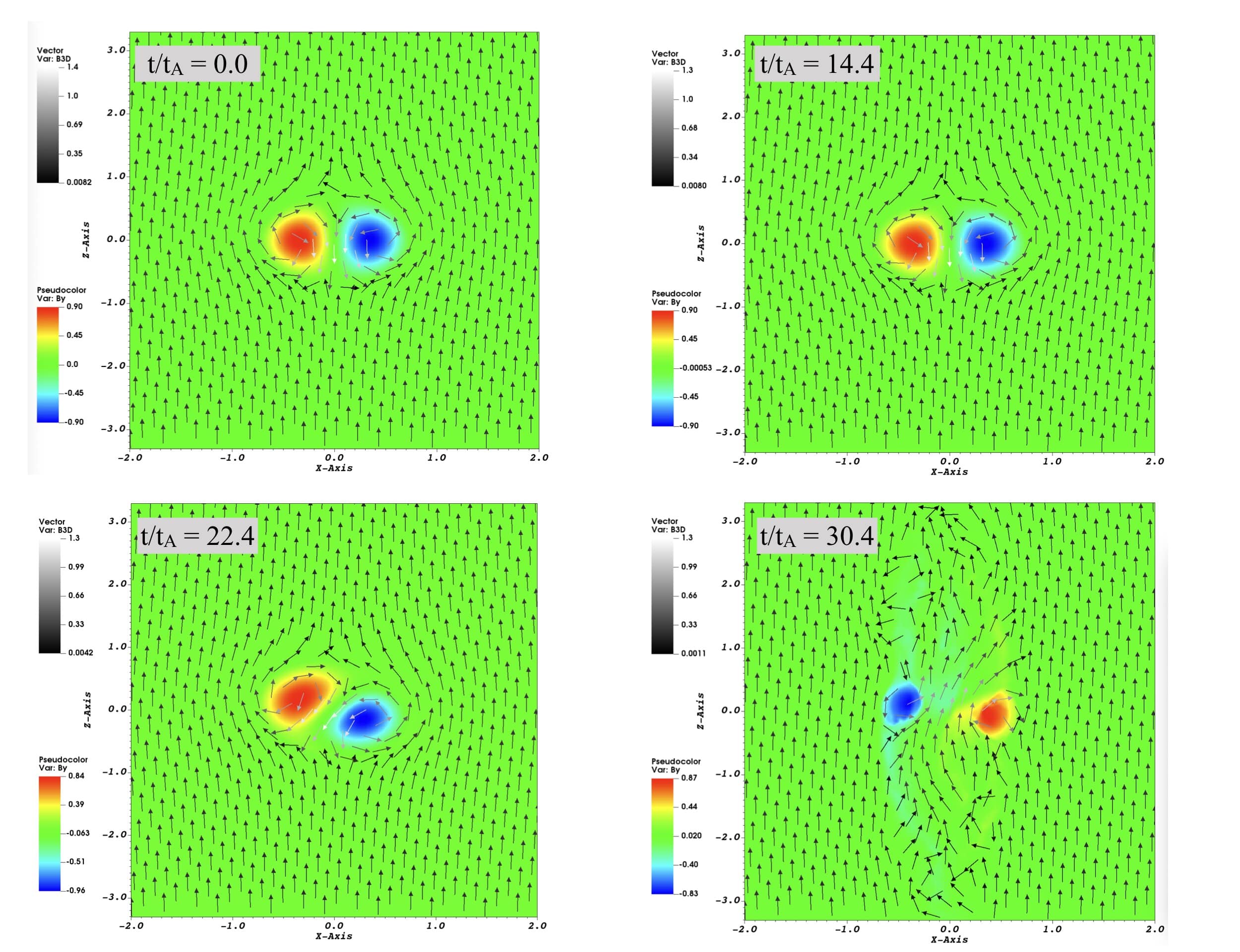}
\caption{
Same as Fig. \protect \ref{fig:sp1z} but showing 
 the value of toroidal magnetic field $B_y$ (color scheme);   vectors depict $\boldsymbol{B3D}$.} 
\label{fig:toroidal}
\end{figure}

\begin{figure}
\centering
\includegraphics[width=\textwidth]{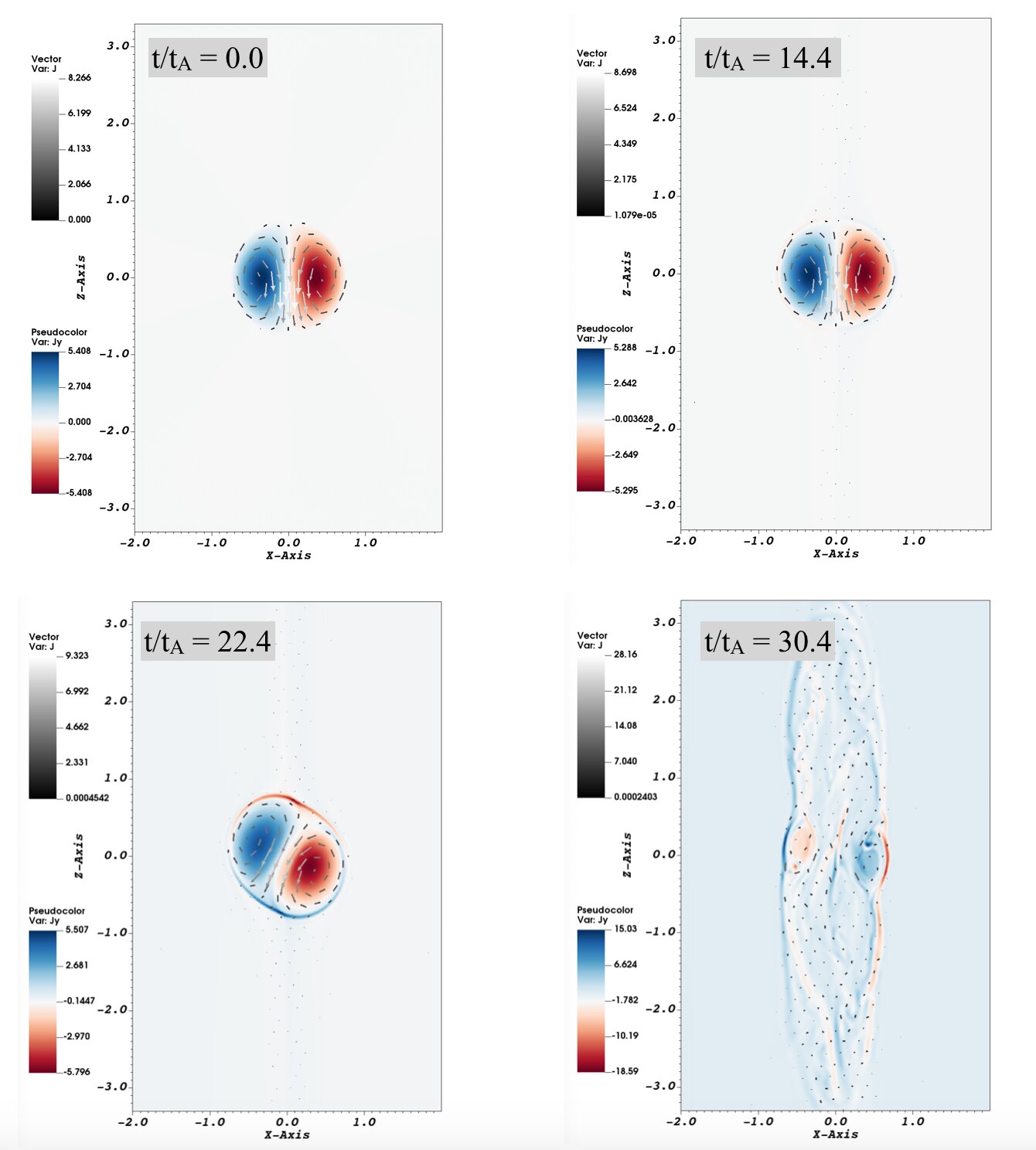}
\caption{Same as Fig. \protect \ref{fig:sp1z}  showing 
 the value of the toroidal current density $J_y$  (color) and  vectors  $\boldsymbol{J}$. The third panel clearly shows the formation of a surface current sheet as the spheromak rotates.} 
\label{fig:current}
\end{figure}

Initially at $t=0$, the constant density plasma is in the relaxed lowest energy state - a spheromak composed of magnetic islands shown by red and blue blobs in \fig{toroidal} symmetrical on either side of the $x=0$ ($z$) axis, depicting poloidal and toroidal components of magnetic field. Magnetic field at the center of spheromak is $-B_0\hat{z}$, that is, spheromak's magnetic moment is anti-aligned with the external magnetic field. A basic spheromak is thus unstable against tilt.

Spheromak begins to tilt immediately after $t=0$. At $t\sim 14.4t_A$, the plasma density inside the spheromak decreases slightly. This is because once dissipation starts, some of the trapped magnetic energy is converted into heat and at the same time magnetic tension within the spheromak decreases. As a result, the spheromak expands and plasma density decreases. At $t\sim 22.4t_A$, tilting is clearly visible; spheromak starts to  rotate about the center and tries to align its magnetic moment  with the external field to lower its energy. As the spheromak tilts, the matching of internal and external \Bfs\  no longer holds, resulting in a current sheet formation on its surface and the onset of magnetic reconnection. The third panel of current density plot in \fig{current} clearly shows the formation of this current sheet on the surface of spheromak. It should be noted that while there is no resistivity in ideal MHD, the process responsible for dissipation, current sheet formation and magnetic reconnection is numerical resistivity arising due to errors introduced by spatial and temporal discretization.

The simulation terminates at $t\sim 30.4t_A$, when plasma hits the walls of the simulation domain. In this quasi-final state, which marks the partial disruption of the spheromak, plasma becomes less dense, magnetic islands rotate fully and magnetic field lines near the center are aligned with the $z$-axis. In \fig{sp1z}, smaller magnetic islands are still seen about the center and magnetic field at their edges is opposite to the external field. Current sheets are still visible around these residual magnetic islands, as seen in the fourth panel of \fig{current}. If the simulation is made to run longer, these magnetic islands will also reconnect at the edges and dissipate. 3D MHD simulation of the lowest energy Taylor state is concurrent with the argument made in \S\hyperref[sec:tilt]{2.2} that a spheromak confined in external magnetic field is intrinsically unstable; it first tries to tilt to lower its energy and eventually dissipates.
\subsubsection{Tilt Instability Growth Rate and Magnetic Energy Dissipation}
\label{sec:tilt}

\begin{figure}
\centering
\begin{subfigure}{0.4\textwidth}
\includegraphics[width=\linewidth]{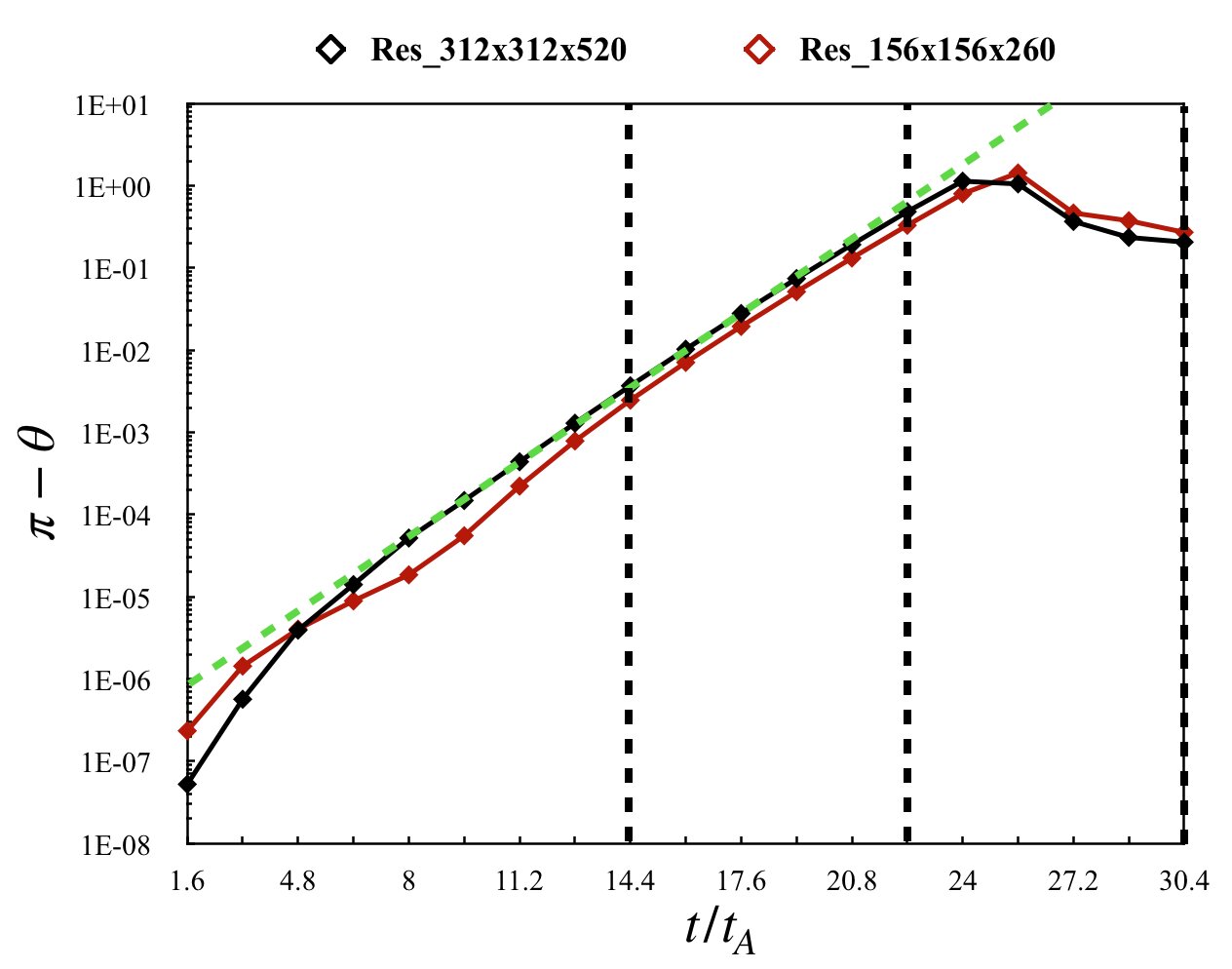}
\caption{}
\end{subfigure}
\begin{subfigure}{0.43\textwidth}
\includegraphics[width=\linewidth]{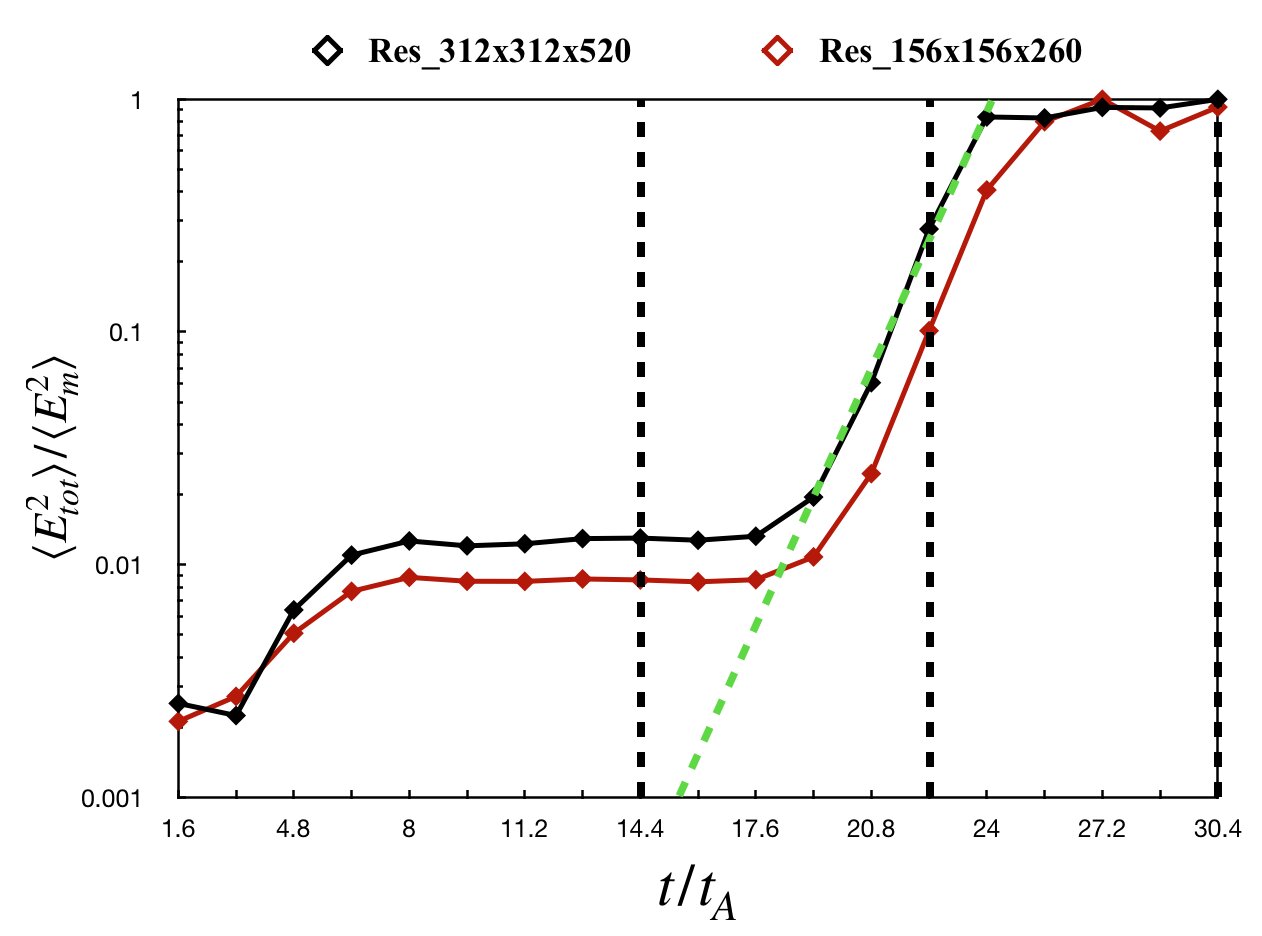}
\caption{}
\end{subfigure}
\begin{subfigure}{0.43\textwidth}
\includegraphics[width=\linewidth]{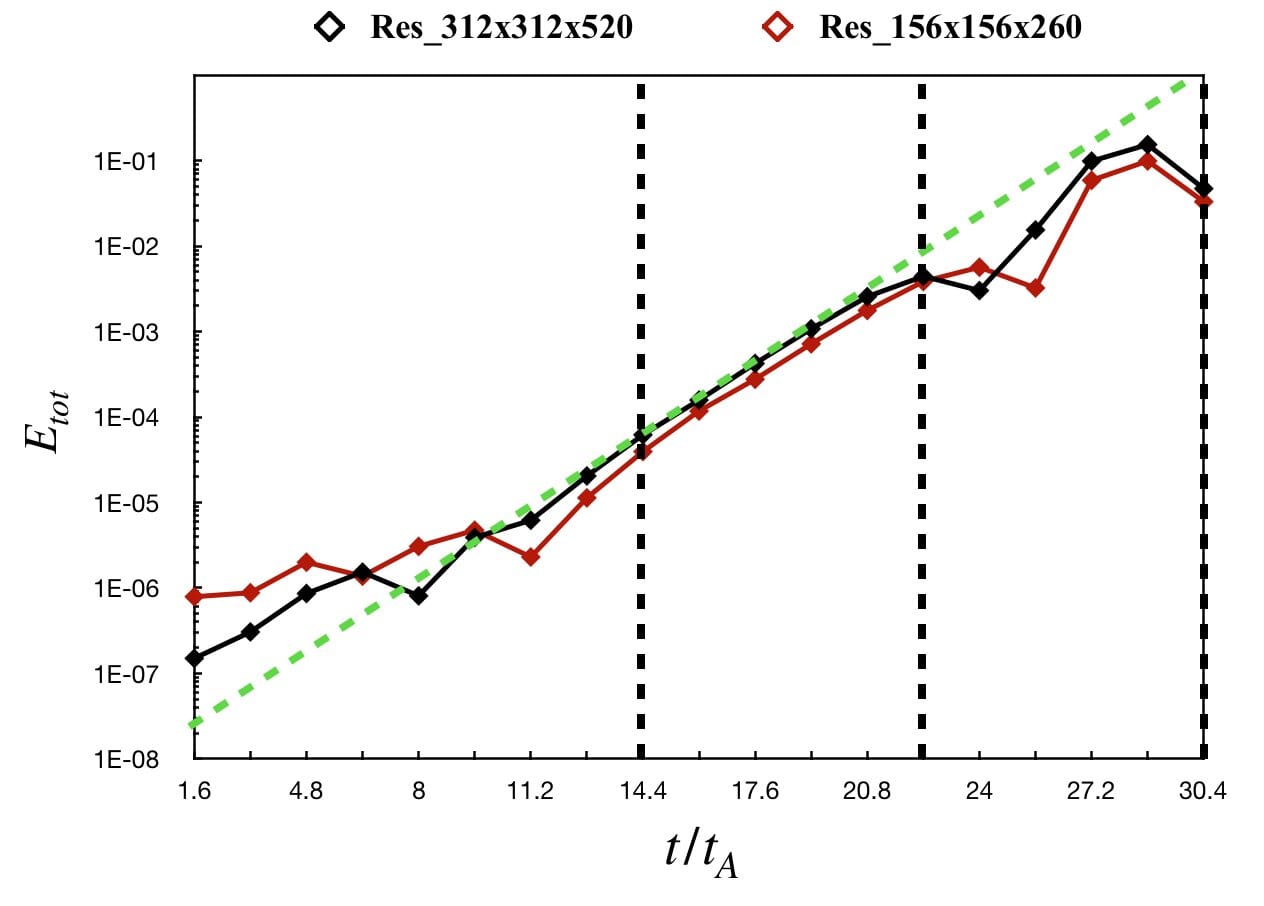}
\caption{}
\end{subfigure}
\caption{(a) Time evolution of the tilt angle $\theta$ in log-linear scale. (b) Time evolution of $\langle E_{tot}^2\rangle/\langle E_{m}^2\rangle$ in log-linear scale. (c) Time evolution of $E_{tot}$ at the center of spheromak in log-linear scale. In all three, a clear phase of exponential growth can be seen (green dotted line). From the plots, $(\pi - \theta)\propto \exp{(0.64v_A t/r_0)}$, $\langle E_{tot}^2\rangle/\langle E_{m}^2\rangle \propto \exp{(0.8v_A t/r_0)}$ and $E_{tot} \propto \exp{(0.6v_A t/r_0)}$. Spheromak dissipates in $\sim 20t_A$ over which instability grows linearly with a growth rate of $0.64/t_A$. Vertical dashed lines indicate the time snapshots used for \fig{sp1z} \& \fig{toroidal}.}
\label{fig:ins}
\end{figure}

\begin{figure}
\centering
\begin{subfigure}{0.45\textwidth}
\includegraphics[width=\linewidth]{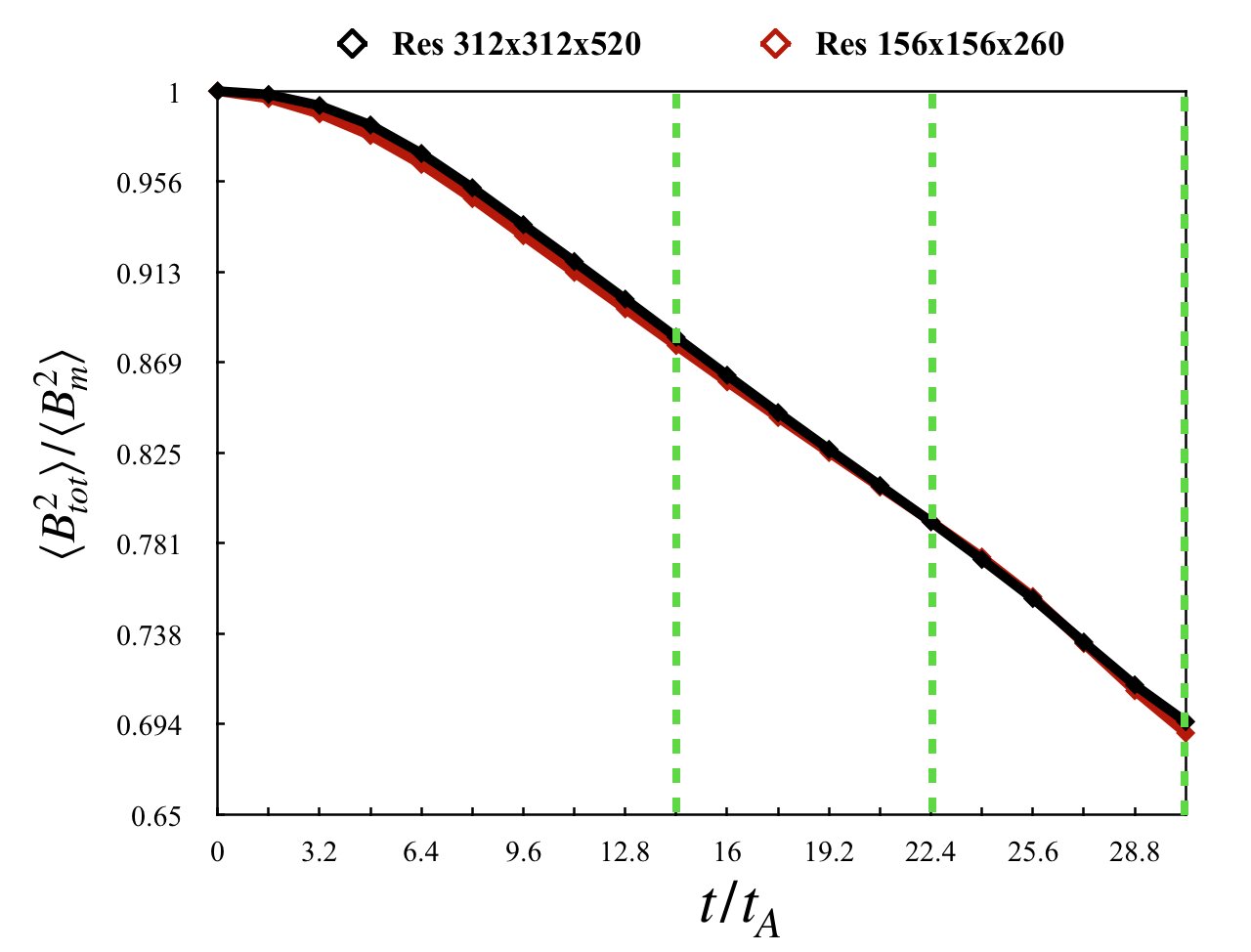}
\caption{}
\end{subfigure}
\begin{subfigure}{0.49\textwidth}
\includegraphics[width=\linewidth]{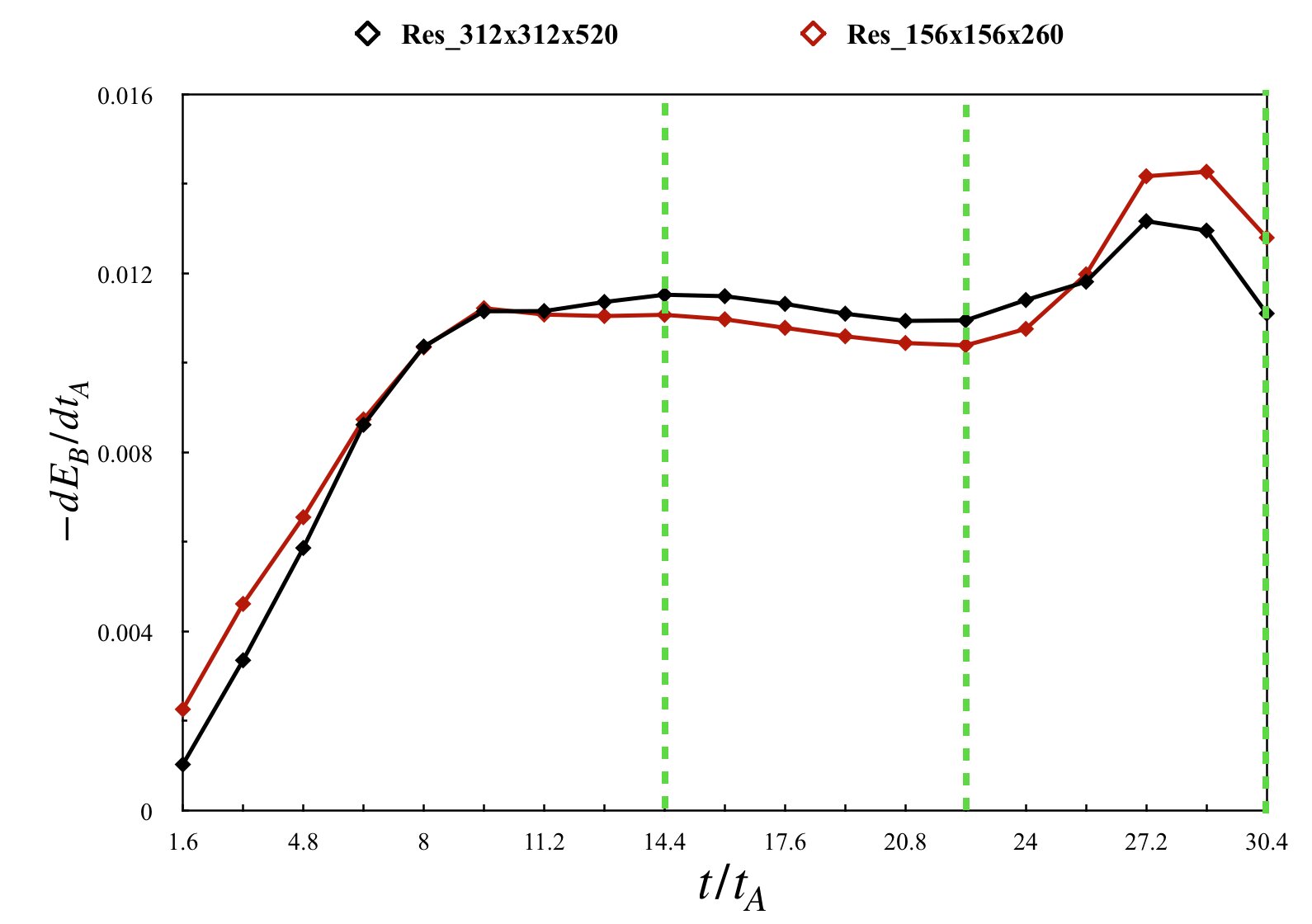}
\caption{}
\end{subfigure}
\caption{(a) Time evolution of box-averaged total magnetic energy for the two different resolutions. Total magnetic energy is plotted in terms of ${\langle B_{tot}^2\rangle}/{\langle B_m^2\rangle}$. About $30\%$ of the initial magnetic energy in the simulation box is dissipated when the spheromak tilts and starts dissipating, eventually hitting the walls. (b) Time evolution of rate of magnetic energy release. Initially, there is a steady increase in the rate until $\simeq 8t_A$ after which magnetic energy is released at a constant rate throughout the duration of tilt instability growth. Green dashed lines indicate the time snapshots used for \fig{sp1z} \& \fig{toroidal}.}
\label{fig:energy_1r}
\end{figure}

\fig{ins} depicts the time evolution of $\theta$, $\langle E_{tot}^2\rangle/\langle E_{m}^2\rangle$ and $E_{tot}$ for the two different resolutions (I) and (II) and also show convergence. Here, $\theta$ is the tilt angle defined as the angle between the total magnetic field at the origin and $z$-axis, $\langle E_{tot}^2\rangle$ is the box averaged squared of total electric field, $\langle E_{m}^2\rangle$ is the maximum value of $\langle E_{tot}^2\rangle$ and $E_{tot}$ is the total electric field at the center of spheromak. We choose to normalize the box-averaged squared of the total electric field by its maximum value in the box assuming the maximum value to remain constant for the duration of the simulation. This helps to visualize the behavior of electric field for both resolutions simultaneously. We also checked the behavior of $y$ component of electric field $E_y$ at the spheromak's center and box-averaged $\langle E_{y}^2\rangle$ and they show the same trend as $E_{tot}$ at the center and box-averaged $\langle E_{tot}^2\rangle$ respectively.

Panel (a) shows an initially anti-aligned spheromak with $\theta \approx \pi$ radians. For simulation (I), a straight line fit to the linear phase of the plot clearly depicts an exponential growth of instability within the spheromak from $t \simeq 5t_A$ to $t \simeq 24t_A$ where $\theta^{'} = (\pi - \theta)\propto \exp{(0.64v_A t/r_0)}$. We can quantify the growth rate of tilting through an angle $\theta^{'}$ by

\begin{equation}
  \gamma_t = \frac{1}{\theta^{'}} \frac{d\theta^{'}}{dt}
  \label{growthrate_gamma}
\end{equation}
giving $\gamma_t = 0.64/t_A$. 

The timescale of dissipation of spheromak is $\sim 20t_A$. Similar fits to the linear phase of plots (b) and (c) give $\langle E_{tot}^2\rangle/\langle E_{m}^2\rangle \propto \exp{(0.8v_A t/r_0)}$ and $E_{tot} \propto \exp{(0.6v_A t/r_0)}$. Here, we use three distinct measures to estimate the instability growth rate and it is seen that they are slightly different but consistent with each other. These results are also in good agreement with the analysis using PIC simulation which will be shown in \S  \ref{sec:lorenzo}.

We also plot the time evolution of box-averaged total magnetic energy in terms of ${\langle B_{tot}^2\rangle}/{\langle B_m^2\rangle}$ and time evolution of rate magnetic energy release for the two different resolutions. Here, ${\langle B_{tot}^2\rangle}$ is the box-averaged squared of total magnetic field, ${\langle B_{m}^2\rangle}$ is the maximum value of ${\langle B_{tot}^2\rangle}$ and $E_B$ is the magnetic energy in the box. \fig{energy_1r} shows that the results are independent of resolution. Panel (a) shows that about $30\%$ of the total magnetic energy is dissipated from the box during the entire evolution of the spheromak from $t=0$ to $t=30.4t_A$. Interestingly, as shown in panel (b), the rate of magnetic energy release stays almost constant during the exponential growth of instability.

For simulation (I), we can also estimate an initial magnetic flux in the $xy$ plane by summing the value of the $z$ component of magnetic field over an $xy$ slice at t=0; we find that it is smaller than the value of $B_0 \times N_x \times N_y$, the total magnetic flux in the box without an embedded spheromak. This is due to the fact that magnetic field lines effectively get pushed out of the simulation box once a spheromak is introduced. Thus, it is not very physical to track the time evolution of excess energy in the box (\eg the difference between the total magnetic energy within the box in the presence of spheromak and energy in the constant magnetic field).

\subsection{Qualitative Picture of Spheromak Instability}

 \begin{figure}
\centering
\includegraphics[width=\textwidth]{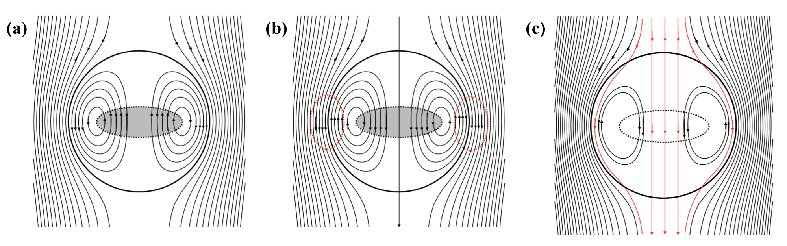}
\caption{Qualitative evolution of tilting instability. Plotted are poloidal magnetic field lines in the $xz$ plane. Initial spheromak (a) is unstable to tilting, so that the spheromak flips over (b), creating current sheets on the surfaces (highlighted in red-dashed). At the core the field inside the spheromak is aligned with external field (gray circle in the center). Reconnection at the surface connects the internal field lines to the external field (c) -  newly reconnected field lines are highlighted in red. At the same time the external field connects to the fields close to the center. At this stage there is a donut-shaped toroidal configuration with still counter aligned fields - this is clearly seen in simulations, last panels in \fig{sp1z} \& \fig{toroidal}.}
\label{fig:reconnection}
\end{figure}

The time evolution of lowest energy Taylor state described in $\S\hyperref[sec:taylor]{3.3.2}$ by 3D MHD simulations can be described qualitatively, see  \fig{reconnection}. Approximately, the spheromak first flips by $180$ degrees, and then reconnects the part of the magnetic flux counter  to the 
external \Bf.

Let us discuss  the  properties of the configuration after the spheromak flips, but before any substantial dissipation sets in.
In the equatorial plane $\theta=\pi/2$ there exists a disk of radius $r_*=2.74/\lambda$, defined by the condition $B_\theta=0$ (it is depicted by the gray circle in the center of  panels (a) and (b) in \fig{reconnection}) within which all field lines point along the external field and whose boundary separates it from the region where the field lines are opposite to the external field. This poloidal magnetic field which is directed opposite to the external field constitutes a poloidal flux $\psi_{opp}$ in the equatorial plane that would eventually reconnect with the external field. We estimate this flux using Eq. (\ref{Bin}) in the following discussion.

The total poloidal flux through the spheromak is zero, $\int_{0}^{r_0}B_\theta 2\pi r dr = 0$, composed of two counter-aligned contributions at $r< r^*$ and $r> r^*$, each of value  
\begin{equation}
 \psi_{opp} = \int_{0}^{r_*}B_\theta 2\pi r dr =2.26 B_0r_0^2
\end{equation}

This is the amount of poloidal flux in the equatorial plane that reconnects and eventually dissipates. Panel (c) in \fig{reconnection} shows partial dissipation of spheromak where newly connected field lines are highlighted in red. At the same time external field connects to the fields close to the center. At this stage there is a donut-shaped toroidal configuration with still counter aligned fields. This is  clearly seen in the last panels of 3D MHD simulations of \fig{sp1z} \& \fig{toroidal}.

\subsection{Evolution of the Second Order Spheromak}
\label{second}

\begin{figure}
\centering
\includegraphics[width=\textwidth]{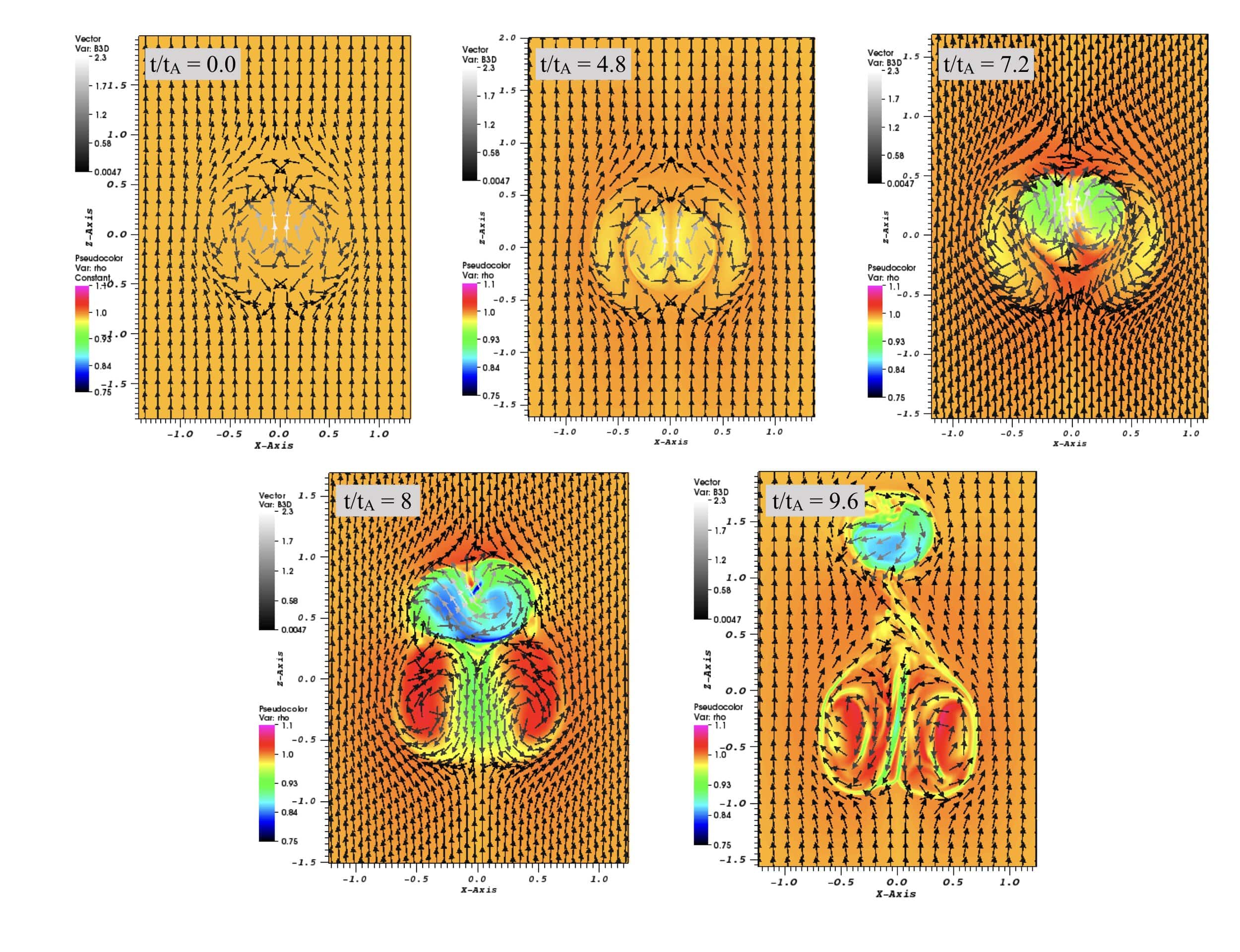}
\caption{Slice in the $xz$ plane of MHD simulation of 2-root spheromak with $\lambda \approx 7.725/{r_0}$. Times are indicated in the panels in units of the Alfv\'{e}nic crossing time $t_A=r_0/v_A$. Colors indicate plasma density while vectors depict $\boldsymbol{B3D}$. The 2-root spheromak goes from being symmetrical to the inner spheromak almost totally detaching from the outer one in $\sim 9.6t_A$.}.
\label{fig:sp2z}
\end{figure}

\begin{figure}
\centering
\begin{subfigure}{0.46\textwidth}
\includegraphics[width=\linewidth]{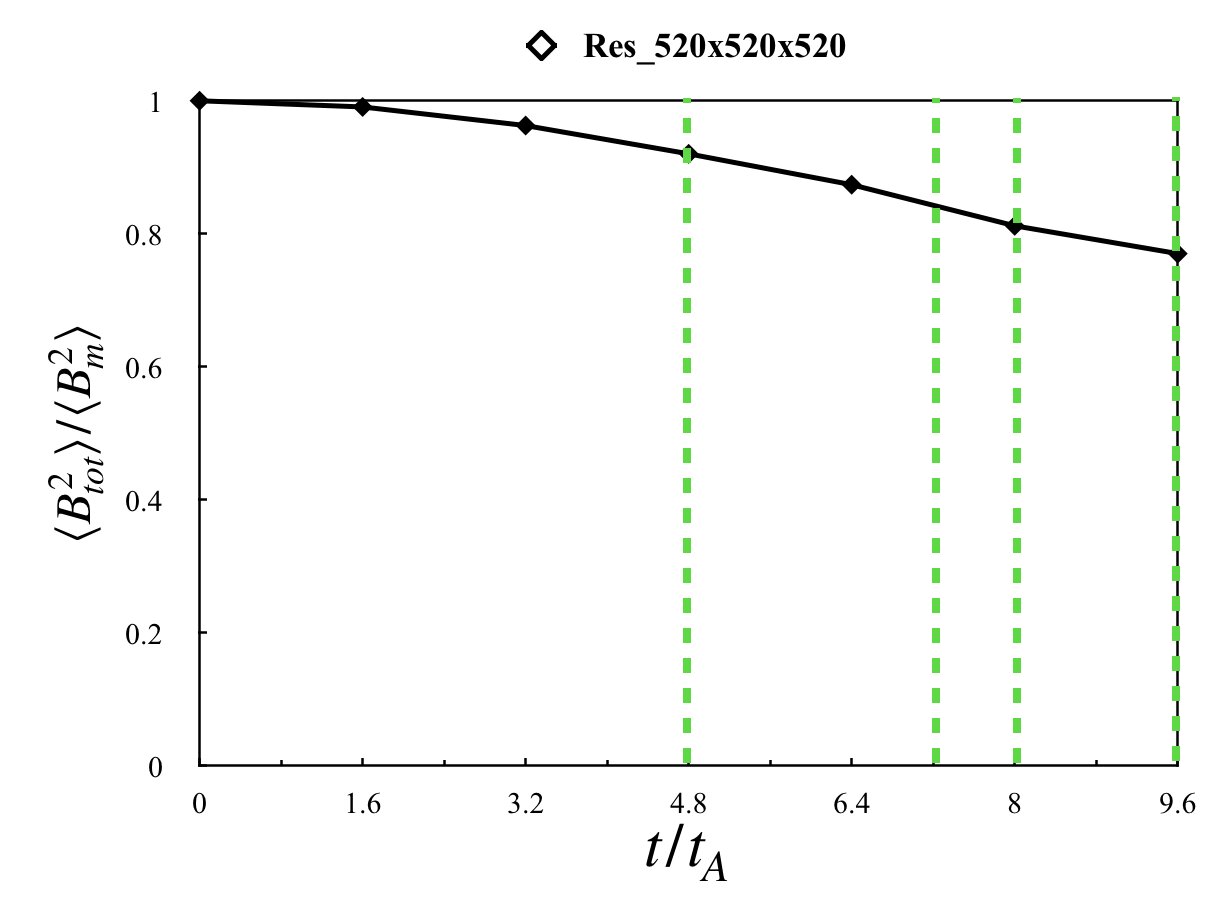}
\caption{}
\end{subfigure}
\begin{subfigure}{0.46\textwidth}
\includegraphics[width=\linewidth]{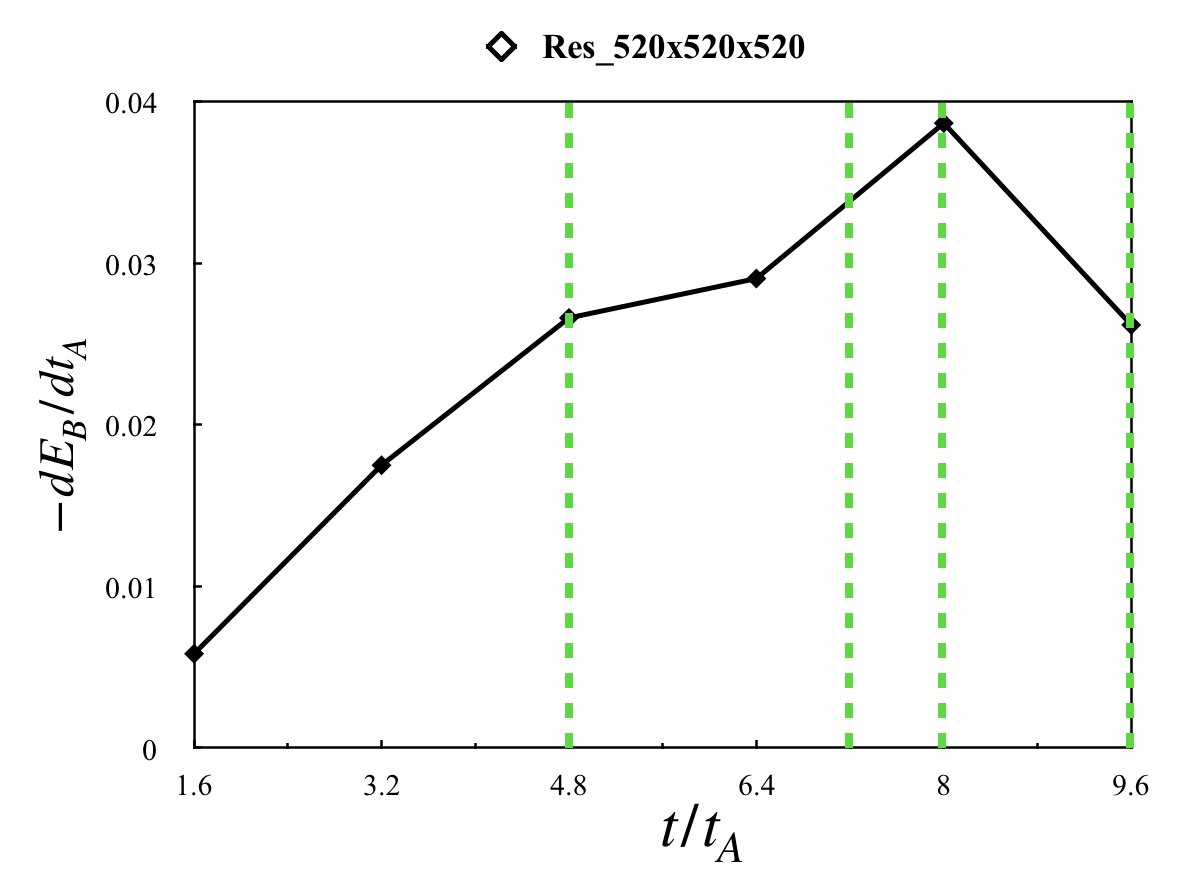}
\caption{}
\end{subfigure}
\caption{(a) Time evolution of box-averaged total magnetic energy in terms of ${\langle B_{tot}^2\rangle}/{\langle B_m^2\rangle}$. About $23\%$ of the initial magnetic energy in the simulation box is dissipated when the 2-root spheromak goes from being symmetrical to the inner spheromak almost totally separating from the outer one. (b) Time evolution of rate of magnetic energy release. There is a gradual increase in the rate throughout the entire evolution. Green dashed lines indicate the time snapshots used for \fig{sp2z}.}
\label{fig:energy_2r}
\end{figure}

In addition to the lowest energy Taylor state, we also simulated  the second order spheromak, corresponding to the   second zero of the spherical Bessel function, $\lambda \approx 7.725/{r_0}$, see  \fig{sp2z}. 
This case can be thought of as  an example of a twisted magnetic configuration (the inner core), confined by another twisted configuration (the outer shell). 

In these  simulations the size of the domain is $x \in [-2, 2]$, $y \in [-2, 2]$ and $z \in [-2, 2]$. Uniform resolution is used in the computational domain with total number of cells $N_{\rm X}=N_{\rm Y}$=$N_{\rm Z}$ = 520. At time $\sim7t_A$, the inner spheromak starts to get expelled from the outer one.
By the time $\sim 9.6 t_A$, the smaller inner spheromak almost totally disconnects from the outer  spheromak; the density within it decreases considerably due to magnetic dissipation. After the expulsion of the inner core the two spheromaks evolve nearly independently, similar to the basic spheromak case considered in \S  \ref{sec:taylor}.


Similar to the basic spheromak, we show the time evolution of box-averaged total magnetic energy in terms of ${\langle B_{tot}^2\rangle}/{\langle B_m^2\rangle}$ and time evolution of rate magnetic energy release in \fig{energy_2r}. Panel (a) shows that about $23\%$ of the total magnetic energy is dissipated from the box during the entire evolution of the 2-root spheromak from $t=0$ to $t=9.6t_A$. Panel (b) depicts that magnetic energy is released at an increasing rate throughout the evolution unlike the basic spheromak where there was a nearly flat phase during the instability growth.

\subsection{PIC Simulation of  Basic Spheromak }
\label{sec:lorenzo}

\begin{figure}
\centering
\includegraphics[width=\textwidth]{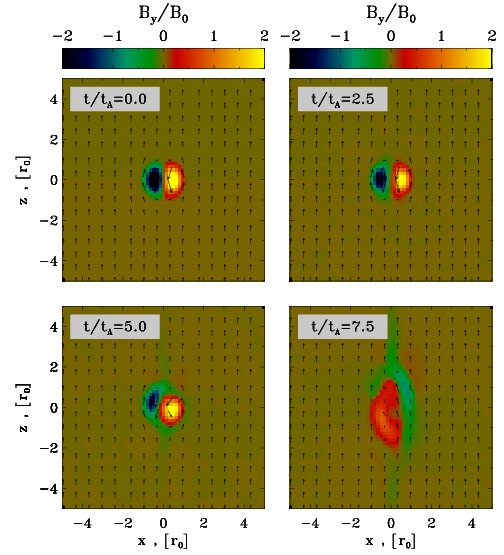}
\caption{PIC simulation of the time evolution of the lowest order Taylor state. Times are indicated in the panels in units of the Alfv\`enic crossing time $t_A=r_0/v_A$. Colors indicate the value of $B_y/B_0$ in the $xz$ plane going through the center of the spheromak, while arrows indicate the $B_x$ and $B_z$ components.}
\label{fig:fluidtime}
\end{figure}

\begin{figure*}
\centering
\includegraphics[width=0.8\textwidth]{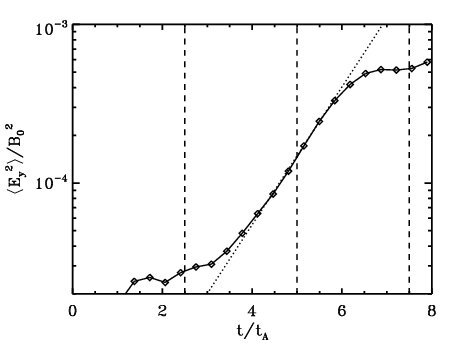}
\caption{From the PIC simulation of the lowest order Taylor state, we show the time evolution of box-averaged $\langle E_y^2\rangle/B_0^2$ in log-linear scale, where $E_y$ is the $y$-component of electric field. Vertical dashed lines indicate the time snapshots used for \fig{fluidtime}. A clear phase of exponential growth can be seen from $t/t_A\simeq 3$ to $t/t_A\simeq 6$, with $\langle E_y^2\rangle \propto \exp{(v_A t/r_0)}$ (dotted line). }
\label{fig:growth}
\end{figure*}

We have supplemented our MHD simulations with particle-in-cell (PIC) simulations performed with the 3D electromagnetic PIC code TRISTAN-MP \citep{buneman_93,spitkovsky_05}. We employ a 3D cube with 1440 cells on each side, and periodic boundary conditions in all directions. The domain is initialized with uniform density of cold electron-positron plasma, with 2 computational particles per cell. The skin depth $\comp$ is resolved with 2.5 cells. The radius $r_0$ of the spheromak is $50\comp=125$ cells. The strength of the magnetic field $B_0$ is calibrated such that the magnetization $\sigma=B_0^2/(4\pi n_0 m c^2)=10$, where $n_0$ is the total particle density, $m$ the electron (or positron) mass and $c$ the speed of light. This implies that the Alfv\'en speed $v_A=c\sqrt{\sigma/(1+\sigma)}\simeq0.95\,c$.

\fig{fluidtime} shows the evolution of the magnetic field $B_y/B_0$ in the $xz$ plane passing through the center of the spheromak. Arrows represent the $B_x$ and $B_z$ components in that plane. The top left panel presents the initial state of the system. At early times (top right panel), the configuration is still close to the initial conditions, while at later times (bottom left) the spheromak starts to tilt, in analogy with the MHD simulations presented above. The final state of the system (bottom right) is also similar to the MHD results.

Further insight into the growth of the tilt instability is presented in \fig{growth}, where we show the evolution of box-averaged $\langle E_y^2\rangle/B_0^2$, where $E_y$ is the $y$-component of electric field. Vertical dashed lines indicate the time snapshots used for \fig{fluidtime}. A clear phase of exponential growth can be seen from $t/t_A\simeq 3$ to $t/t_A\simeq 6$, with an $\langle E_y^2\rangle$ growth rate $\simeq v_A/r_0$ (dotted line). We have checked that the growth rate scales as $r_0^{-1}$ by performing a similar simulation with $r_0=75\comp$. The measured growth rate of the instability is in agreement with that estimated from MHD simulations.


\section{Discussion and Conclusions}
\label{sec:5}

In this paper we consider the tilting instability of magnetically confined spheromaks using 3D MHD and PIC simulations. We consider astrophysically important  mildly relativistic regime, when the \Alfven\  velocity  approaches the velocity of light. In addition to basic spheromak  \citep{Ripperda2017} we also consider a second order spheromak, as an example of a magnetically twisted configuration (the inner core) confined by the magnetically twisted shell. 

We find that in all cases  confined spheromak are highly unstable to tilting instabilities.  The instability is driven by the fact that initially the magnetic moment of the spheromak is counter-aligned with the confining magnetic field. As a result the spheromak flips, indicative of a tilt instability. This creates current layers at the boundary. The resulting reconnection  between internal and confining \Bf\ leads to partial annihilation of the spheromak's  poloidal magnetic flux with the external \Bf. 
At the same time the toroidal \Bf\ and the associated helicity (or relative helicity \citep{1994PPCF...36..945J,Bellan2018}) of the initial configuration is carried away by torsional \Alfven waves (in a sense that initial configuration had finite helicity, while the eventual final configuration - just straight magnetic field lines - has zero helicity). 

The evolution of the basic spheromak is generally consistent with previous results. 
The tilting instability of spheromak in cylindrical geometry has been explored by \cite{Bondeson1981} and \cite{Finn1981} where they analyze growth rate of tilting as a function of elongation $L/R$ (see Fig. 4 in both) and derive a threshold value $L/R \approx 1.67$. For our case, $L/R = 2$ and growth rate of $0.64/t_A$ is consistent with the growth rates implied from their Fig. 4 namely $\sim 0.1/t_A$ \citep{Bondeson1981} and $\sim 10/t_A$ \citep{Finn1981}.
An experimental identification of tilting mode of spheromak plasma and its control is discussed in \cite{Munson1985}. A clear exponential growth rate of tilting is visible in their Fig. 1 and is strikingly similar to our \fig{ins} (a).

A characteristic timescale of tilting instability is $\simeq20{t_A}$ during which the spheromak dissipates after losing a significant fraction of its energy which is in good agreement with \cite{Hayashi1983} where they study spheromak dynamics for a force-free plasma by a 3D MHD code and estimate a growth rate of the order of $10t_A$. Interestingly, their results also show that the tilting angle saturates at $90^{\circ}$ unlike our results where the spheromak almost entirely undergoes a $180^{\circ}$ rotation - it flips. The $90^{\circ}$ tilt stabilization of \cite{Hayashi1983} is facilitated by a cylindrical vacuum vessel - a toroidal flux core having a small enough aspect ratio so that further tilting is energetically unfavorable \citep{Bellan2018}. A similar characteristic growth time of tilt around the magnetic axis and use of a flux conserver to stabilize the tilt mode is suggested in \cite{Jarboe1994} which also provides an excellent review on formation and stability of spheromaks. 


We have also studied the evolution of second order magnetically confined spheromak as an example of a configuration (the inner core) confined by the twisted \Bf\ (the outer shell). Very quickly ($\sim10t_A$) the inner core separates from the outer shell and  completely detaches. As a result two nearly independent dissipative structures are formed. No stabilization occurs.

Our results disfavor models of magnetically  confined  structures for the origin of  tail oscillations in  magnetar flares (\cite{Lyutikov2003,2008MNRAS.387.1735M}), as we discuss next.
Magnetars are young ($\sim 10^3 - 10^4$ years) and highly magnetized (surface magnetic fields $\sim 10^{14} - 10^{15}$ G) neutron stars exhibiting  X-ray and $\gamma$-ray activity. Most dramatic giant flare till date was exhibited by SGR 1806-20 on December 27, 2004 \citep{Palmer2005,Mereg2005} in which the main spike that  lasted $\sim0.5$ seconds was  followed by a $\sim380$ s pulsating tail. This   is  $\sim50$ cycles of high-amplitude pulsations at the SGR's known rotation period of 7.56 s.   The long pulsating tails of giant flares originate in ``trapped fireball''  that remains confined to the star's closed magnetic field lines. 

In magnetar's \mss\ the Alfv{\'e}n speed through a plasma of density $\rho$ is nearly relativistic  \citep{Gedalin1993}:

\begin{equation}
 \frac{v_A}{c} = \left(\frac{B^2/4\pi}{\epsilon+P+B^2/4\pi}\right) ^{1/2} \approx 1
\end{equation}
$c$ is the speed of light, $\epsilon=\rho c^2$ is the total energy density of plasma particles and $P$ is the total plasma pressure. For a magnetically dominated plasma, $P, \, \epsilon << B^2/4\pi$.
Thus, the Alfv{\'e}n time within the magnetar's \ms\
\begin{equation}
t_{A}=\frac{R_{NS}}{v_{A}} \approx 3 \times 10^{-5}  \text{s} 
\end{equation}
where $R_{NS}=10$ km is the radius of a \NS. Our results demonstrate that stabilization  even of higher order spheromaks does not occur, so that 
the timescale over which a spheromak confined in the magnetar's magnetosphere would dissipate is  too short to explain the tail duration
\begin{equation}
t_{\rm diss} \sim 20t_{A} \approx  \times 10^{-3}  \text{s}
\end{equation}

Finally, let us comment on the applicability of the Taylor relaxation principle to astrophysical plasmas. It was suggested in \cite{Bellan2000} that spheromak is a Taylor state, so that evolution of the system will lead to the largest possible spheromak.
The Taylor relaxation principle assumes that the plasma is surrounded by a wall  impenetrable to helicity escape. This can be achieved in a laboratory, with arrangements of conducting walls. This is not possible in astrophysical surrounding as we argue next.

First, according to the  Shafranov's virial theorem \citep[eg.][]{Bellanbook}   it is not possible to have an isolated self-contained MHD equilibrium  -  there must always be some external confining structure. It is possible to have purely unmagnetized external confining structures - one can construct spheromak-type configurations confined by external pressure  \cite{2010MNRAS.409.1660G}. The configurations considered  by \cite{2010MNRAS.409.1660G} are not  force-free, but they look very similar to spheromaks. They are stable to current-driven instabilities. 
It seems the case considered by \cite{2010MNRAS.409.1660G} is the only case when Taylor relaxation principle would be applicable to astrophysical plasmas -  if there is non-zero B-field in the confining medium the spheromak will try to flip and  will reconnect.  This will generally happen very fast, on few \Alfven time scale.  The helicity will then be emitted as Alfven shear waves; this then violates the Taylor principle  of conserved helicity. 

Thus, astrophysical magnetic configurations belong more naturally to a class called ``driven magnetic configurations'' by \cite{Bellan2018} - they are generally  magnetically  connected to some outside medium. As a result of this connection helicity will leave the system in the form of torsional Alfven waves. This will violate the assumptions of Taylor relaxation scheme. 


We explore a possible astrophysical application of our numerical results. Using the energetics of SGR 1806-20, the estimated dissipation timescale of a magnetically confined spheromak is of the order of a milli second, whereas the quasi-periodic oscillations in the SGR's giant flare release energy for $\sim400$ s. The formation and spontaneous dissipation of a spheromak in a magnetar's magnetosphere doesn't allow for such prolonged energy release. It would be worthwhile to explore coalescence instability in turbulent plasmas. It has been suggested in \cite{Reiman1982} that by Taylor's theory, repeated coalescence of $n$ spheromaks of equal size increases the radius of the spheromak by a factor of $n^{1/4}$ whereas the total magnetic energy of the final spheromak will be $n^{-1/4}$ times the sum of the energies of the initial spheromaks. We speculate that such a mechanism might stabilize the spheromak over longer timescales. Another important investigation would be to look for effects of plasma rotation on the tilt mode stability in the context of a spheromak using arguments similar to those made in \cite{Mohri1980}, \cite{Ishida1988} and \cite{Ji1998} in which it is shown that plasma rotation in the $\theta$ direction can help stabilize the tilt mode, but in field-reversed configurations (FRCs). Finally, it would serve useful to explore if both, coalescence and rotation together could have stabilizing effects to sustain a spheromak over longer timescales.

\acknowledgements 
\section*{Acknowledgements}
We thank the anonymous referee for their constructive comments which greatly improved the manuscript. The numerical simulations were carried out in the CFCA cluster of National Astronomical Observatory of Japan. We thank the \textit{PLUTO} team for the possibility to use the \textit{PLUTO} code and for technical support. The visualization of the results were performed in the VisIt package. We would like to thank Paul Bellan, Eric Blackman and Oliver Porth for discussions and comments on the manuscript.

This work had been supported by NASA grant 80NSSC17K0757 and NSF grants 1903332 and 1908590. Lorenzo Sironi acknowledges support from the Sloan Fellowship, the Cottrell Scholars Award, and NSF PHY-1903412.

\bibliographystyle{jpp}
\bibliography{spheromak,blob,BibTex}

\end{document}